%% file: sparta.tex
\documentclass[acmsmall]{acmart}

\AtBeginDocument{%
  \providecommand\BibTeX{{%
    \normalfont B\kern-0.5em{\scshape i\kern-0.25em b}\kern-0.8em\TeX}}}

\usepackage{booktabs}
\usepackage{multirow}
\usepackage{comment}
\usepackage{subfig}
\usepackage{graphicx}
\usepackage{algorithm}
\usepackage[noend]{algpseudocode}

\setcopyright{none}

\renewcommand\footnotetextcopyrightpermission[1]{} 
\begin{document}

\title{SPARTA: A Divide and Conquer Approach to Address Translation for Accelerators}


\author{Javier Picorel}
\affiliation{%
  \institution{Huawei Technologies}
  \city{Munich}
  \country{Germany}}
\email{javier.picorel@huawei.com}

\author{Seyed Alireza Sanaee Kohroudi}
\affiliation{%
  \institution{National University of Singapore}
  \country{Singapore}}
\email{sanaee@comp.nus.edu.sg}

\author{Zi Yan}
\affiliation{%
  \institution{NVIDIA Research}
  \country{USA}}
\email{ziy@nvidia.com}

\author{Abhishek Bhattacharjee}
\affiliation{%
  \institution{Yale University}
  \country{USA}}
\email{abhishek.bhattacharjee@yale.edu}

\author{Babak Falsafi}
\affiliation{%
  \institution{\'Ecole Polytechnique F\'ed\'erale de Lausanne}
  \city{Lausanne}
  \country{Switzerland}}
\email{babak.falsafi@epfl.ch}

\author{Djordje Jevdjic}
\affiliation{%
  \institution{National University of Singapore}
  \country{Singapore}}
\email{jevdjic@.nus.edu.sg}

\renewcommand{\shortauthors}{Picorel, et al.}



\input{tex/abstract.tex}

\maketitle
\thispagestyle{empty}

\input{tex/introduction.tex}
\input{tex/background.tex}
\input{tex/our_approach.tex}
\input{tex/hw_support}

\input{tex/os}

\input{tex/methodology}

\input{tex/new_evaluation}

\input{tex/relatedwork}

\input{tex/conclusion}

\bibliographystyle{ACM-Reference-Format}
\bibliography{references}

\end{document}

%% file: tex/abstract.tex
\begin{abstract}

Virtual memory (VM) is critical to the usability and programmability of hardware accelerators. Unfortunately, implementing accelerator VM efficiently is challenging because the area and power constraints make it difficult to employ the large multi-level TLBs used in general-purpose CPUs. Recent research proposals advocate a number of restrictions on virtual-to-physical address mappings in order to reduce the TLB size or increase its reach. However, such restrictions are unattractive because they forgo many of the original benefits of traditional VM, such as demand paging and copy-on-write.

We propose SPARTA, a divide and conquer approach to address translation. SPARTA splits the address translation into accelerator-side and memory-side parts. The accelerator-side translation hardware consists of a tiny TLB covering only the accelerator's cache hierarchy (if any), while the translation for main memory accesses is performed by shared memory-side TLBs. Performing the translation for memory accesses on the memory side allows SPARTA to overlap data fetch with translation, and avoids the replication of TLB entries for data shared among accelerators. To further improve the performance and efficiency of the memory-side translation, SPARTA logically partitions the memory space, delegating translation to small and efficient per-partition translation hardware. Our evaluation on index-traversal accelerators shows that SPARTA virtually eliminates translation overhead, reducing it by over 30x on average (up to 47x) and improving performance by 57\%. At the same time, SPARTA requires minimal accelerator-side translation hardware, reduces the total number of TLB entries in the system, gracefully scales with memory size, and preserves all key VM functionalities.

\end{abstract}

%% file: tex/introduction.tex
\section{Introduction}
\label{sec:intro}

\begin{figure*}
\centering
 \includegraphics[width=1\textwidth,clip]{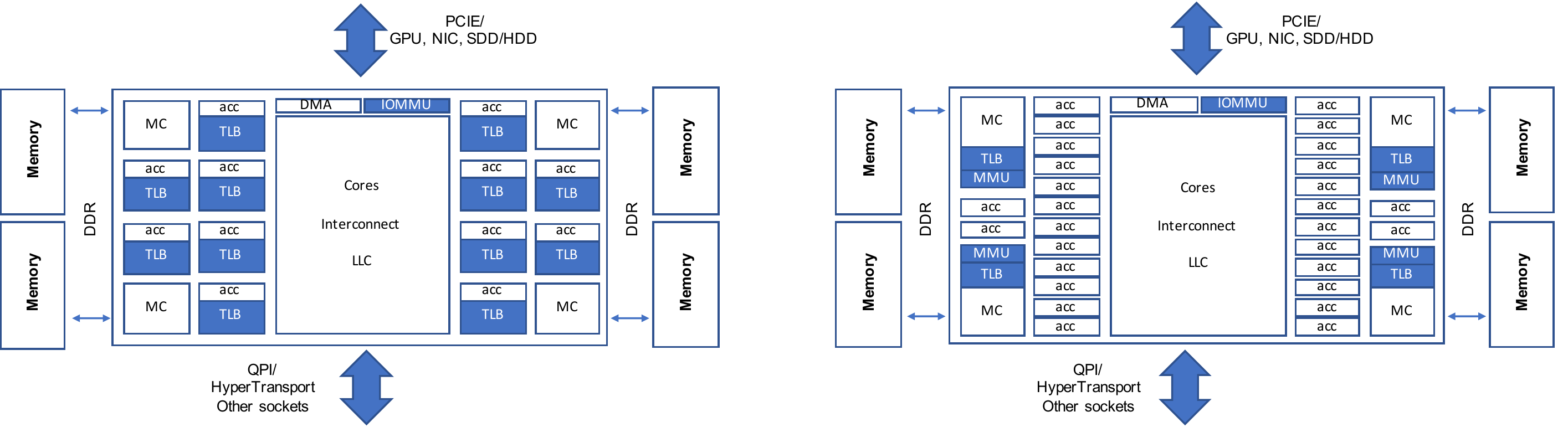}
 \caption{Overview of conventional (left) and SPARTA (right) systems. SPARTA uses IOMMU only for legacy accelerators.}
\label{fig:overview}
\end{figure*}

Architecting virtual memory (VM) for hardware accelerators has become a first-order
goal for chip designers and operating system developers
\cite{pichai:architectural, power:supporting, haria:devirtualizing, shin:scheduling}. There are many
reasons for this including the fact that VM and global address space programming ease software development effort by enabling ``a pointer is a pointer everywhere'' semantics \cite{pichai:architectural,
vesely:observation, ausavarungnirun:mosaic}, extending memory protection to accelerators, and obviating
the need for manual CPU-accelerator data marshalling. 

Unfortunately, it is challenging to extend VM to
accelerators efficiently. General-purpose CPUs (and GPUs) use
large hierarchical TLBs to achieve satisfactory VM performance, but such
large structures are ill-suited to emerging area- and power-constrained
accelerators for neural networks and deep learning \cite{reagen:minerva}, graph processing \cite{haria:devirtualizing},
and in/near memory processing \cite{picorel:near-memory}. Moreover, accelerators
rely on a centralized IOMMU for page table walks \cite{olson:border}. However,
IOMMUs are physically distant from both accelerators and the increasingly large and distributed
memory, making walks even higher-latency events on accelerators than on CPUs \cite{shin:scheduling}.

Recent studies suggest approaches that encourage or enforce specific mapping functions between virtual 
and physical addresses~\cite{basu:efficient, gandhi:range, haria:devirtualizing, picorel:near-memory}. This permits
custom TLB hardware to compress translations to increase the TLB reach and/or reduce its size~\cite{pham:colt, pham:increasing, basu:efficient,haria:devirtualizing}. 
While superpages were early examples of these efforts \cite{navarro:practical, cox:efficient},
recent work uses serendipitously-generated translation contiguity in virtual
and physical spaces \cite{pham:colt, pham:increasing, cox:efficient} to compress TLB entries. 
More aggressive segment-based approaches~\cite{basu:efficient, gandhi:range} map hundreds of consecutive 
virtual pages to consecutive physical frames, while identity-mapping approaches~\cite{haria:devirtualizing}
go even further by requiring identical virtual and physical addresses. 
These approaches enable extreme TLB compression, but preclude many key benefits of 
VM, such as copy-on-write (CoW) and demand paging~\cite{basu:efficient, haria:devirtualizing}, require the entire dataset to fit in memory, and can be hard to realize in real-world settings where memory becomes highly 
fragmented because of memory-intensive and long-running applications or multi-tenancy. 
All of the above approaches require significant changes to TLB hardware to support different compression techniques. 
Direct-mapped/set-associative VM approaches, such as DIPTA~\cite{picorel:near-memory} treat memory as a set-associative virtual cache,
allowing a virtual page to map to a single physical frame (or a couple of frames). In doing so, DIPTA largely overlaps translation and
data fetch, but its hardware-based page table incurs prohibitive overhead of 2MB of SRAM for every GB of memory and introduces intrusive changes to DRAM. Being a virtual cache, it suffers from the synonym problem, does not support CoW, and its low associativity may 
significantly increase the page-fault rate, particularly for multi-programmed workloads~\cite{picorel:near-memory}. It is also worth noting that extreme mapping restrictions that many of the above mechanisms impose significantly reduce the entropy in the virtual-to-physical address mapping, creating a potential security vulnerability~\cite{haria:devirtualizing}.

In this paper we seek to address the inefficiencies and inflexibility of prior work by proposing a novel address translation mechanism called SPARTA (Split and PARtitioned Translation for Accelerators). SPARTA splits the task of address translation between the accelerators and the memory in order to reduce the overall performance and hardware overheads. The accelerator-side translation hardware consists of a tiny TLB whose reach covers only the accelerator's cache hierarchy (if any). As a result, cache hits enjoy fast address translation at minimal area and power costs, which is particularly attractive for accelerators, while translation for cache misses is performed by shared memory-side TLBs/MMUs. Our decision to limit the reach of accelerator-side TLBs is based on the observation that memory-side translation could be largely overlapped with data fetch if both memory-side translation hardware and the relevant page-table entries are in sufficient proximity to the data that needs to be fetched. Under such circumstances, accelerator-side TLB hits provide no performance benefit upon cache misses. To maximize the overlap between data fetch and memory-side translation, SPARTA divides the physical memory space into logical partitions (of arbitrary size) and ensures that a virtual address uniquely identifies the partition holding the data. Once the target partition is identified, SPARTA issues data fetch and translation in parallel, performing translation on the memory side using a local per-partition TLB/MMU situated next to the corresponding partition. To further minimize page walk latency in case of memory-side TLB misses, SPARTA's design ensures that data and the corresponding page-table entries are always residing in the same partition, providing highly localized page-table walks.

Apart from virtually eliminating the translation overhead, SPARTA achieves a number of important goals: \raisebox{.5pt}{\textcircled{\raisebox{-.9pt} {1}}} It imposes minimal restrictions on the virtual-to-physical mapping. Unlike identity-mappings~\cite{haria:devirtualizing} or DIPTA~\cite{picorel:near-memory}, which provide one~\cite{haria:devirtualizing} or a couple of options~\cite{picorel:near-memory} for page placement, SPARTA allows a page to be mapped anywhere within the partition (millions of options). \raisebox{.5pt}{\textcircled{\raisebox{-.9pt} {2}}} Unlike segment-based approaches~\cite{basu:efficient, gandhi:range}, SPARTA requires no contiguity in virtual-to-physical mapping, as consecutive virtual pages do not have to be mapped to consecutive physical pages. Consequently, SPARTA works well with arbitrary levels of memory fragmentation. \raisebox{.5pt}{\textcircled{\raisebox{-.9pt} {3}}} SPARTA supports key VM functionalities, such as demand paging and CoW, it does not suffer from the synonym problem and does not require the datasets to fit in memory. \raisebox{.5pt}{\textcircled{\raisebox{-.9pt} {4}}} SPARTA requires minimal translation hardware on the accelerator-side, which makes it suitable for arbitrarily small accelerators. \raisebox{.5pt}{\textcircled{\raisebox{-.9pt} {5}}} Because per-partition TLBs are responsible for disjoint parts of the physical address space and because TLB entries are shared among the accelerators, the replication of TLB entries for data that are shared among accelerators is avoided. Consequently, SPARTA significantly reduces the total number of TLB entries in the system. \raisebox{.5pt}{\textcircled{\raisebox{-.9pt} {6}}} Because each MMU's TLB hierarchy serves only its local memory partition, SPARTA enables graceful performance scaling across memory chip/partition counts. \raisebox{.5pt}{\textcircled{\raisebox{-.9pt} {7}}}
 SPARTA minimally reduces the entropy in virtual-to-physical mapping, avoiding the security vulnerabilities of other approaches. 

We compare SPARTA with well-known translation mechanisms. With modest memory-side TLBs, SPARTA reduces translation overhead 
by 31.5x on average (up to 47x) over conventional translation, and by 19x over huge pages (up to 28x), improving performance by 57\% on average (43\% over huge pages). To identify the key parts of production OSes that require changes to support SPARTA, we also prototype SPARTA in stock Linux. The fact that our small team of researchers was able to integrate these kernel changes in a manageable
number of man-hours lends credence to the readily-implementable nature of SPARTA's software requirements.

%% file: tex/background.tex
\section{Background and Motivation}
\label{sec:background}


\begin{figure}[t]
    \begin{minipage}{0.70\textwidth}
    \centering
   \includegraphics[width=1.0\textwidth]{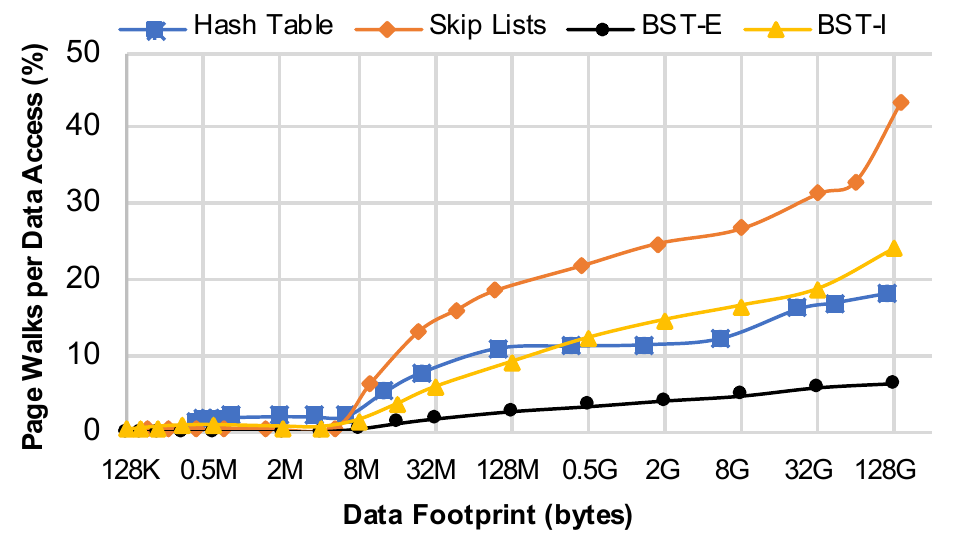}
   \caption{Page walk rate as a function of memory size.}
   \label{fig:pagewalks}
   \end{minipage}
\end{figure}

\begin{table*}
\centering
\caption{Comparison of SPARTA with previous VM approaches. CoW stands for copy-on-write. Protection stands the ability to provide fine-grained per-page protection. Security indicates high entropy in virtual address bits.}
\label{table:vms}
\resizebox{\linewidth}{!}{
\begin{tabular}{l|c|c|c|c|c|c|c|c|}
\cline{2-9}
                                                                                             & \multicolumn{1}{l|}{Programmability} & \multicolumn{1}{l|}{Performance} & \multicolumn{1}{l|}{Efficiency} & \multicolumn{1}{l|}{\begin{tabular}[c]{@{}l@{}}Demand Paging\end{tabular}} & \multicolumn{1}{l|}{CoW} & \multicolumn{1}{l|}{\begin{tabular}[c]{@{}l@{}}Robust to Fragmentaion\end{tabular}} & \multicolumn{1}{l|}{Protection} & \multicolumn{1}{l|}{Security} \\ \hline

\multicolumn{1}{|l|}{Multi-page mappings \cite{pham:colt, pham:increasing}} & {\color[HTML]{009901} yes}             & {\color[HTML]{FE0000} no}          & {\color[HTML]{FE0000} no}         & {\color[HTML]{009901} yes}                                                      & {\color[HTML]{009901} yes} & {\color[HTML]{009901} yes}                                                              & {\color[HTML]{009901} yes}        & {\color[HTML]{009901} yes}      \\ \hline
\multicolumn{1}{|l|}{Transparent Huge Pages \cite{transparenthugepages}}    & {\color[HTML]{009901} yes}             & {\color[HTML]{FE0000} no}          & {\color[HTML]{FE0000} no}         & {\color[HTML]{009901} yes}                                                      & {\color[HTML]{009901} yes} & {\color[HTML]{FE0000} no}                                                               & {\color[HTML]{FE0000} no}         & {\color[HTML]{009901} yes}      \\ \hline
\multicolumn{1}{|l|}{Libhugetlbfs \cite{lighugetlbfs}}                      & {\color[HTML]{FE0000} no}              & {\color[HTML]{FE0000} no}          & {\color[HTML]{FE0000} no}         & {\color[HTML]{009901} yes}                                                      & {\color[HTML]{009901} yes} & {\color[HTML]{FE0000} no}                                                               & {\color[HTML]{FE0000} no}         & {\color[HTML]{009901} yes}      \\ \hline
\multicolumn{1}{|l|}{Direct Segments \cite{basu:efficient}}                 & {\color[HTML]{FE0000} no}              & {\color[HTML]{009901} yes}         & {\color[HTML]{009901} yes}        & {\color[HTML]{FE0000} no}                                                       & {\color[HTML]{FE0000} no}  & {\color[HTML]{FE0000} no}                                                               & {\color[HTML]{FE0000} no}         & {\color[HTML]{FE0000} no}       \\ \hline
\multicolumn{1}{|l|}{Redundant Memory Mappings \cite{karakostas:redundant}} & {\color[HTML]{009901} yes}             & {\color[HTML]{FE0000} no}          & {\color[HTML]{FE0000} no}         & {\color[HTML]{FE0000} no}                                                       & {\color[HTML]{FE0000} no}  & {\color[HTML]{FE0000} no}                                                               & {\color[HTML]{FE0000} no}         & {\color[HTML]{FE0000} no}       \\ \hline
\multicolumn{1}{|l|}{Identity Mappings \cite{vesely:observation}}           & {\color[HTML]{009901} yes}             & {\color[HTML]{009901} yes}         & {\color[HTML]{009901} yes}        & {\color[HTML]{FE0000} no}                                                       & {\color[HTML]{FE0000} no}  & {\color[HTML]{FE0000} no}                                                               & {\color[HTML]{FE0000} no}         & {\color[HTML]{FE0000} no}       \\ \hline
\multicolumn{1}{|l|}{Set-Associative VM \cite{picorel:near-memory}}         & {\color[HTML]{009901} yes}             & {\color[HTML]{009901} yes}         & {\color[HTML]{FE0000} no}         & {\color[HTML]{009901} yes}                                                      & {\color[HTML]{FE0000} no}  & {\color[HTML]{009901} yes}                                                              & {\color[HTML]{009901} yes}        & {\color[HTML]{FE0000} no}       \\ \hline
\multicolumn{1}{|l|}{SPARTA (this work)}                                                     & {\color[HTML]{009901} yes}             & {\color[HTML]{009901} yes}         & {\color[HTML]{009901} yes}        & {\color[HTML]{009901} yes}                                                      & {\color[HTML]{009901} yes} & {\color[HTML]{009901} yes}                                                              & {\color[HTML]{009901} yes}        & {\color[HTML]{009901} yes}      \\ \hline
\end{tabular}
}
\end{table*}

\subsection{Goals for Large Heterogeneous Systems}
Similar to prior work~\cite{haria:devirtualizing}, we identify the following as key goals for our design:

\begin{itemize}
        \item \textbf{Programmability.} Unified virtual memory between CPUs and
          accelerators simplifies data sharing and eliminates
          hand-managed data copying and marshalling. Ideally, we wish
          to preserve all the benefits typically associated with VM
          like memory protection and isolation, and flexibility of
          sharing parts of the address space among processes.

        \item \textbf{Flexibility.} Conventional VM imposes few
          restrictions on virtual-to-physical page mapping. 
          This flexibility is valuable to perform allocations
          when memory is fragmented, to support application
          multi-tenancy, enable memory allocation strategies transparent to
          programmers, to support applications that do not fit in memory, and to integrate key features such as
          demand paging and copy-on-write (CoW). Recent proposals reduce or preclude many of these
	  cababilities\cite{haria:devirtualizing, basu:efficient, picorel:near-memory,karakostas:redundant,lighugetlbfs}. 

        \item \textbf{Safety and Security.} Direct access to physical
          memory is generally not desirable. Such
          memory management approaches cannot prevent malicious or
          erroneous memory accesses and prohibit sharing accelerators
          across different processes with proper
          isolation~\cite{haria:devirtualizing}. Furthermore, the
          entropy in address mappings must be as high as possible to
          reduce vulnerability to security attacks.

        \item \textbf{Performance and Efficiency.} The
          goals listed thus far must be achieved efficiently
          without excessive on-chip area overheads. In other words,
          address translation must provide near-zero performance
          overheads regardless of the application working set and locality
          patterns, and must do so under tight area and power
          constraints (particularly for area- and power-constrained accelerators).

\end{itemize}

\subsection{Shortcomings of Modern MMU Hardware}
The ever-increasing memory needs of modern software have given rise to
scale-out systems with large memories that provide low-latency access
to data \cite{ferdman:clearing, karakostas:performance, volos:fat,
  basu:efficient}. Larger memory sizes are
problematic for both address translation reach and latency, as we discuss next.

\subsubsection{TLB Reach.}
It is difficult to build TLBs
with sufficient capacity or {\it reach} to cover increasing physical
memory sizes \cite{basu:efficient, haria:devirtualizing,
  pham:colt}. Chip vendors are aggressively growing TLBs
-- e.g., Intel has been doubling CPU TLBs from Sandybridge to Skylake chips -- and paying the cost of increased
area/power. But despite TLBs with thousands of entries,
poor locality of memory references in emerging big-memory workloads elevates TLB miss
rates. Figure \ref{fig:pagewalks} captures this effect by
quantifying TLB miss rates as we vary the memory footprint of several
of our workloads on an Intel Broadwell chip with 1.5K-entry L2 TLBs
(see Section \ref{sec:methodology} for details). Despite Broadwell's
large TLBs, miss rates increase dramatically with
memory footprint, corroborating results from prior work
\cite{basu:efficient}. 

Such problems are pernicious on accelerators too. For example,
GPUs integrate massive multi-thousand-entry TLBs
too \cite{vesely:observation, lowepower:inferring}, but there is
widespread consensus that this approach is not viable for more
area-constrained accelerators \cite{haria:devirtualizing,
  picorel:near-memory}. While techniques like large pages can offer partial relief in some cases,
they present their own set of challenges because they offer only
coarse-grained protection \cite{pham:large}, can be hard to form on
fragmented systems \cite{kwon:coordinated}, have poor NUMA support
\cite{gaud:large}, and require complex TLB hardware for concurrent
page size support \cite{cox:efficient}. For these reasons,
transparent support for large pages in OSes like Linux apply only to
2MB pages (and not other sizes like 1GB) even after decades of
research \cite{arcangeli:transparent}. Practically, vendors implement
multi-thousand-entry TLBs for the worst-case scenario when base 4KB
pages dominate. We concur with recent
work \cite{pham:colt, basu:efficient, karakostas:redundant,
  haria:devirtualizing} that while large pages should certainly be improved
and can be effective, other orthogonal approaches (like ours) are
needed to operate in a complementary manner, as shown in Section~\ref{sec:evaluation}.

\subsubsection{Page Table Walk Latency.} 
The penalty of a TLB miss is also critical to
address translation performance. CPUs are equipped
with dedicated MMU caches to accelerate page table walks
\cite{bhattacharjee:large-reach, barr:translation}. Unfortunately,
even with MMU caches, page table walk latencies can be unacceptably high
for accelerators. The key culprit is that
heterogeneous systems with accelerators are usually integrated with
NUMA memories and require increasingly long-latency lookups across sockets/chips,
etc. Even perfect MMU caches require at least one memory reference per
page table walk and recent work shows that this single reference can
dramatically exacerbate address translation costs for accelerators
\cite{picorel:near-memory, pichai:architectural}.

\subsection{Prior Approaches}

Table \ref{table:vms} summarizes prior work on improving address translation. 
While we include superpages, their shortcomings spurred the proposal
of all the other techniques included in the table. Unfortunately, these alternatives change VM software 
and sacrifice aspects of traditional VM flexibility (to varying degrees) to 
achieve effective address translation. SPARTA's goal is to find a better 
compromise between retaining traditional VM benefits at the software level, 
while also enabling more efficient TLB hardware.

\vspace{2mm}
\noindent\textbf{Multi-page mappings.} COLT~\cite{pham:colt} and clustered~\cite{pham:increasing}
TLBs coalesce 4-8 page translations into a single TLB entry, as long
as their physical locations are contiguous. Although TLB reach
improves, they cannot cover the entirety of a large memory system of
tens or hundreds of GBs~\cite{gandhi:range}.  Equally problematically,
these techniques rely on contiguity that {\it might} be
serendipitously generated but does not have to be. Achieving such
contiguity is challenging in highly-loaded cloud systems where memory
can be fragmented \cite{pham:colt}.

\vspace{2mm}
\noindent\textbf{Segments.} Variable-sized
segments can be used instead of fixed page-based
translations~\cite{basu:efficient, karakostas:redundant, park:hybrid} but require invasive  changes to the OS's allocation path to
generate contiguity at memory allocation (i.e., eager paging), preclude the use of CoW and per-page permissions. 
Direct segments require programmers to explicitly allocate a main segment at startup~\cite{basu:efficient}, 
while redundant memory mappings (RMMs)~\cite{karakostas:redundant} 
dynamically manage multiple ranges. The reliance on at-allocation contiguity means that there
are situations on highly fragmented systems where segments/ranges may
be difficult to generate. Additionally, they require highly associative area- and
power-hungry TLBs to take advantage of large ranges of contiguous translations
they generate.

\vspace{2mm}
\noindent\textbf{Direct mappings.} Devirtualized
virtual memory extends the previous line of work on segments
for accelerators
\cite{haria:devirtualizing}. The idea is to identity map virtual
and physical pages for large portions of the application footprint,
enabling extreme TLB information compression for the identity mapping. 
While address translation overheads are largely eliminated, identity mappings are optimistic in
highly-loaded multi-tenant cloud scenarios with memory
fragmentation. Similar restrictions apply to other direct-mapped/extremely low-associativity VM
approaches (like DIPTA \cite{picorel:near-memory}) proposed for near-memory accelerators. Importantly, these 
approaches preclude mechanisms like CoW, which are widely used in fork
system calls, and their impact on schemes that require address bit
entropy (like ASLR) require further examination from the security aspect before widespread
adoption. Furthermore, approaches like DIPTA require tens to hundres of MBs of SRAM storage or 
intrusive DRAM changes~\cite{picorel:near-memory}.

%% file: tex/our_approach.tex
\section{Our Approach}
\label{our-approach}
Prior approaches have to trade off flexibility in the virtual-to-physical 
mapping for more performance or area- and power-efficient translation hardware. 
Our approach, SPARTA, breaks this tradeoff by introducing minimal restrictions 
to the address mapping that do not preclude the most important benefits of VM,
yet manage to almost entirely overlap memory-side translation with data fetch.
This is in contrast to identity-mappings~\cite{haria:devirtualizing} and 
set-associative VM~\cite{picorel:near-memory}, which also largely overlap translation
and data fetch, but introduce severe mapping restrictions, sacrificing many of the key VM
benefits.

\vspace{2mm}
{\noindent \bf Basic idea:} Figure~\ref{fig:overview} illustrates the idea behind SPARTA.
A virtual page is by design guaranteed to map to a physical page in a 
specific memory ``partition'', but can reside anywhere within the
partition. This partition can be a socket or a memory channel or a further memory division. 
For accelerators without caches, the accelerator-side translation simply comes down to dispatching
the virtual address to an appropriate memory-partition. For accelerators with physical caches,
the accelerator-side translation includes a tiny TLB that only covers the cache (omitted from Figure~\ref{fig:overview} for clarity).
Partitioning permits replacement of a large accelerator TLB that needs to address the
entire physical address space with group of shared memory-side TLBs responsible 
for holding translations only within the local memory partition. Because partitions
represent just a small subset of the total physical address space,
memory-side TLBs can achieve high hit rates despite being smaller in
size. In contrast to accelerator-side TLBs, memory-side TLBs are shared among 
accelerators. As a result, translations corresponding to shared data are never 
replicated, unlike the accelerator-side TLBs. As a result, in the presence of
data sharing, the required overall number of TLB entries in the system is significantly lower compared to the
baseline. When combined with an adequate page-table implementation, partitioning also 
allows us to dramatically reduce memory-side TLB miss penalty by allowing co-location 
of data and the corresponding translations in the same memory partition, 
overlapping their lookup. 

\vspace{2mm}
{\noindent \bf Feasibility and implications on the programming model:}
Implementing memory partitioning in a commodity OS
requires only modest changes to existing memory allocation code-paths, which we
discuss in Section~\ref{sec:os}. SPARTA preserves the paged organization
and fine-grained memory protection. The fact that SPARTA does not rely on
contiguity also means that it can continue to operate efficiently in
fragmented memory settings where translation contiguity or identity
mappings may be hard to generate. For the same reason, SPARTA supports demaind paging 
and can be readily used for applications do not fit in memory.
Due to the minimal restriction in virtual-to-address mapping, SPARTA effectively 
does not compromise entropy in virtual address bits. Furthermore, 
imposing only a modest restriction on virtual-to-physical mapping
means that optimizations like copy-on-write can, for all practical purposes, 
be supported (Section~\ref{sec:os}).

%% file: tex/hw_support.tex
\section{SPARTA: Hardware Support}
\label{sec:hw_support}

In this section, we compare the hardware requirements of conventional 
memory management with SPARTA. 

\subsection{Baseline Address Translation}
Figure~\ref{fig:timeline}a and Figure~\ref{fig:timeline}b illustrate
the timeline of conventional address translation. For clarity, we first 
consider the simplest case, which is accelerators without caches (we will later focus on accelerators with caches).
 In the conventional case, the accelerator probes its local TLB. On a TLB hit
(Figure~\ref{fig:timeline}a), the translation completes and
data fetch commences. Because data can be anywhere in memory, NoC and off-chip network traversals 
are usually necessary to reach the target memory controller. DRAM is then accesssed
and data is routed back through the same networks.

On a TLB miss (Figure~\ref{fig:timeline}b), the page table walker determines the physical
address of the page table and initiates its lookup. Page tables
can be implemented in many ways, e.g., radix tree, inverted, hashed,
etc. Consider the best-case scenario where either because of the page
table structure or due to the presence of perfect MMU caches, the
entire page table walk amounts to a single memory access
\cite{bhattacharjee:large-reach, barr:translation}. This access can be
to any memory location. Consequently, the access may require 
traversing the NoCs and off-chip interconnects to reach the memory channel
and memory chip hosting the desired page table entry (PTE). The memory controller then reads the PTE from
DRAM and returns it via the NoCs and off-chip interconnects to the accelerator, where the TLB is filled. 
This part of the timeline constitutes the translation
path. After translation, the accelerator issues the data
fetch. The NoC and off-chip interconnects are traversed to reach
the memory chip that holds the requested data. Once the data is read
from DRAM, the response traverses the same NoC and off-chip
interconnects to reach the accelerator. The part of the
timeline constitutes data fetch.

As Figure~\ref{fig:timeline}b shows, conventional address
translation is expensive. The TLB's accelerator-side
integration means that it must address all of physical memory, which
elevates its miss rate. Furthermore, translation and data paths cannot overlap
and must be serialized. Note further that accelerators are usually not equipped 
with hardware page-table walkers, but instead
ask the IOMMU to perform the walk. Hence, the
miss penalty is inflated with IOMMU access latencies.

\subsection{SPARTA Address Translation}
Due to SPARTA's memory partitioning, each virtual page
maps to a physical frame in a specific memory partition. Therefore, 
the accelerator does not perform complete translation and instead
determines the memory partition that the virtual page resides in. 
The accelerators use a simple hash function (e.g., as simple as using
a subset of the address bits) to route the request to the target partition. 
Per-partition TLBs hold mappings only for physical frames within that partition. 
We can also place a page table within the memory partition only for physical pages belonging
to that partition. Per-partition TLBs are much smaller than
per-accelerator TLBs, which have to cover all of physical
memory, while per-partition page tables reduce page table walk latency as
they are co-located in the same partition as the data.

\begin{figure*}
	\includegraphics[width=\textwidth]{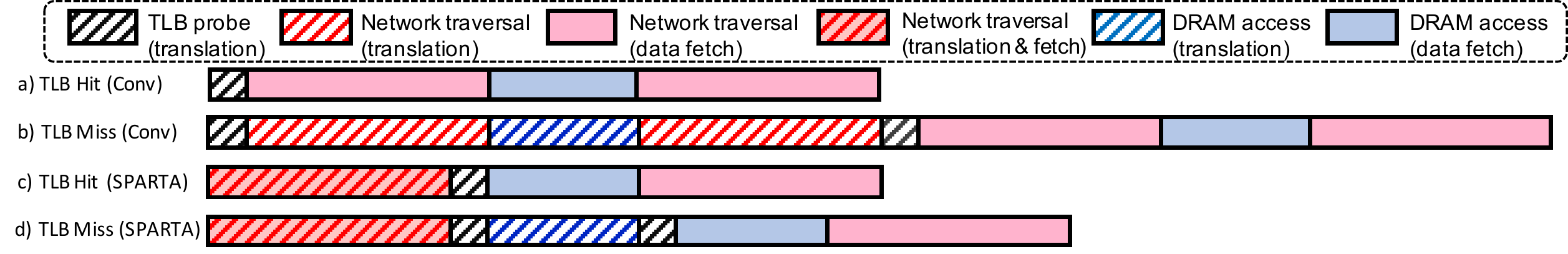}
	\caption{Timeline of events of an accelerator's memory reference in conventional translation and SPARTA (not drawn to scale).}
	\label{fig:timeline}
\end{figure*}

Figure~\ref{fig:timeline}c shows the timeline of events
for a TLB hit in SPARTA. The virtual address uniquely identifies a
memory partition and traverses the network to reach the target memory channel. 
There, a simple multiplexer (not shown in the figure) distinguishes virtual and physical address
requests, and sends the former to the memory partition's TLB, which
completes the translation path and data fetch commences. 
Importantly, SPARTA overlaps translation with data fetch because
both the TLB and data are co-located in the same partition. Although 
translation itself takes longer time, the overhead of translation in case of a TLB hit 
amounts to the TLB probe latency, as in the conventional case.

With SPARTA, TLB misses occur much less frequently. When they do, as
shown in Figure~\ref{fig:timeline}d, the virtual address
identifies the target memory partition, which is reached after traversing the network.
The key difference is that the PTE also resides in the same memory partition. 
Therefore, unlike conventional translation, there is no need
to separately traverse the NoC and off-chip interconnects for page table
lookups and data fetch. Once the PTE arrives from the local DRAM partition and the
TLB fill operation completes, translation finishes. Then, the memory
controller performs a local DRAM access to fetch the data and sends it back
to the accelerator, finishing the data path.

SPARTA page walks are relatively cheap as
the time to locate the target memory channel is part of both the
translation path and data path (essentially removing this part of the
cost of address translation). Page table walks do not involve network
communication as the page table entries are all located in the same
memory channel. The longer the latency of the network is with respect
to the DRAM accesses, the lower overhead of page table walks in SPARTA
relative to the conventional architecture. This behavior particularly
benefits large memory systems, as the need for scaling out the memory
for large-memory machines increases the average distance to the memory
channels, as well as systems with lower DRAM latencies (e.g.,
3D-stacked memories).

Finally, SPARTA's memory-side translation hardware could conceptually
be looked at as a fully distributed IOMMUs. The key difference is that
IOMMUs serve translation/TLB miss requests for the devices attached to
the local socket, but in doing so they cache translations and perform
page walks across the entire memory system. In contrast,  SPARTA's translations
and page walks are confined within the local memory partition, which
is essential scalability. As shown in Figure
\ref{fig:overview}, we anticipate using SPARTA in conjunction with a
standard monolithic IOMMU in platforms where conventional accelerators
like GPUs (which are as close to the memory side as our
accelerators) may need access to them. 

\subsection{Interactions with Caches}
SPARTA can be configured to support accelerators without 
caches~\cite{haria:devirtualizing, picorel:near-memory} (as explained above), but 
also with caches, regardless of whether they are virtually- or physically-addressed. Virtual caches
are increasingly attractive in emerging systems~\cite{yoon:revisiting, park:efficient, shan:legoos} and 
are a natural fit for SPARTA. With virtual caches, SPARTA does not require any translation
 hardware on the accelerator because data in the cache is accessed 
using virtual addresses. Address translation for data absent from the cache is overlapped 
with data fetch. Note that the baseline system with virtual caches still 
requires a sizeable TLB and well-performing MMU to avoid costly address 
translation upon cache misses, whereas such hardware is of little use to SPARTA.

For physical caches, the baseline again requires sizeable TLBs and MMUs on the 
accelerator. SPARTA does also need a TLB, but a tiny one that can cover cache hits 
suffices (detailed analysis in Section~\ref{sec:evaluation}). Because cache 
misses require data fetches from memory and SPARTA effectively overlaps 
these with translation, TLBs with coverage beyond cache hits 
are of little benefit to SPARTA. Note that upon a miss in the accelerator's TLB, 
SPARTA returns both the PTE, which is then cached in the accelerator's TLB, and the requested 
data from memory, which would likely cause a miss in the cache.

%% file: tex/os.tex
\section{SPARTA: Operating System Support}
\label{sec:os}

SPARTA is readily implementable and
compatible with commodity OSes. To showcase that, we have
prototyped SPARTA in stock Linux (4.10) by implementing
configurable memory partitioning in the memory management sections of the
kernel. To co-locate data and page-table entries in the same partition, we 
have implemented SPARTA's page table in the kernel as an inverted page 
table, although other implementations are also possible. These and other ancillary
modifications are fully compatible with the existing kernel. To quantify our 
kernel changes, we run \verb|git diff| and find
the following: \texttt{20 files changed, 266 insertions(+), 43
  deletions(-)}. Key aspects of our changes involved:

\vspace{1mm}
\noindent {\bf Partitioning using NUMA page allocation.} We were
able to simplify our partitioning implementation by leveraging
existing Linux code built to bind processes to NUMA nodes (i.e., {\it
  membind}). We observe that the notion of mapping
only a fixed subset of physical frames to a virtual page is analogous to
the idea of assigning physical frames from only select NUMA
nodes. Thus, we conceptually treat subsets of the virtual address
space as being bound to specific NUMA nodes and implement SPARTA's memory 
partitioning by reusing the {\it membind} code paths. Note
that this implementation would not be possible for set-associative 
VM (i.e., DIPTA ~\cite{picorel:near-memory}), as it would require millions of 
NUMA nodes, which is why DIPTA is bound to hardware page-table implementations.

\vspace{1mm}
\noindent {\bf Virtual page and physical page allocation.} When
implementing SPARTA in Linux, we needed to ensure that each virtual
page maps to a physical page in the right memory partition based on our customizable
{\it MEM\_PARTITION\_INDEX\_HASH()} function. Two situations are possible:
1) a virtual page maps to a physical page that is not present in the
memory, which mostly happens when this physical page is
\textit{private} and a subsequent page fault will allocate a physical
page to this virtual page; 2) a virtual page maps to a physical page
that is already present in the memory, which means either this
physical page is \textit{shared} between processes via the virtual
page or the virtual page is a \textit{remap} of this physical page.
Our solution is to adjust virtual pages or physical pages according to
the existing memory information. If no physical page is present in the
memory, we allocate physical pages according to the memory partition index
based on virtual addresses. If a physical page is present, we assign a
virtual address that maps to the partition which the existing
physical page resides in, based on our indexing scheme. Concrete examples are discussed below:

\begin{itemize}

\item \textbf{Private physical pages.} In this case, because no
    physical page is in the memory before any page fault, we place no
    restriction on virtual address assignment.  This means we use
    Linux’s existing \verb|get_unmap_area|() kernel function to get a
    free virtual address region and assign it to a user space
    application whenever it requests a virtual address space region via malloc. 
    On a page fault, since there is no corresponding physical page in the
    memory, we first calculate the memory partition from the faulting
    virtual address and allocate a physical page, using Linux's NUMA
    page allocation function, from that partition. This
    is shown in Algorithm~\ref{alg:private}.
  
\item \textbf{Shared physical pages.} In this case, because a
    physical page is already present in a specific partition, we have
    to restrict our virtual address assignment. This means that after we
    get a free virtual address region upon a request from user space
    applications by using the kernel function \verb|get_unmap_area|(),
    we need to adjust the starting and ending addresses of the virtual
    address region, so that all the virtual pages in the region map to
    the existing physical pages and conform to our {\it
      MEM\_PARTITION\_INDEX\_HASH()} function. For example, assume a system with
    4 memory partitions, whose physical page regions are [P0...P9],
    [P10...P19], [P20...P29], [P30...P39], and our partition index
    hash function is VPN mod 4.  At the beginning, process Proc1
    maps its virtual memory region [$V3_{Proc1}$ ... $V7_{Proc1}$] to
    a group of physical pages [P30, P0, P10, P20, P31] from 
    partitions 3, 0, 1, 2, 3, respectively.  Next, a second process Proc2
    calls \verb|mmap|() to share the same set of physical pages with
    P1. At that moment, the \verb|get_unmap_area|() kernel function
    can give [$V5_{Proc2}$ ... $V9_{Proc2}$] to P2 as its virtual
    memory region pointing to the shared physical pages. But this
    virtual memory region maps to a sequence of physical pages from
    partitions 1, 2, 3, 0, 1, which does not match the partition
    sequence of the existing physical pages. In that case, inside the
    kernel, we need to adjust the region from
    [$V5_{Proc2}...V9_{Proc2}$] to [$V7_{Proc2}...V11_{Proc2}$] or
    other analogous regions, so that the adjusted virtual memory
    region maps to a sequence of physical pages from partitions 3, 0,
    1, 2, 3, which is the same sequence as the shared physical pages.

\item \textbf{Remapped physical pages.} This case is almost the same as the shared memory case, 
  since it basically assigns a new virtual address region to an existing set of physical pages. 
  Thus, we will first find a suitable virtual memory region, then adjust the virtual address region
  that it maps to the same partition sequence of the group of existing physical pages and return
  it to the userspace application, which asks for memory remapping via \verb|mremap|() system call.
  
\end{itemize}

\noindent {\bf Inverted page table design.} We adopt the same inverted
page design as the hashed page tables in IBM Power systems. We reserve
memory for each inverted page table in each individual NUMA node and
insert/invalidate/update inverted page table entries whenever CPU page
tables change in the same hook functions
(e.g., \verb|update_mmu_cache()|). To guarantee coherence in
translation information between CPUs and accelerators, each inverted
page table entry has a valid bit atomically accessed by both CPUs (for
inverted page table modification) and accelerators (for
reading). 

\vspace{1mm}
\noindent {\bf Summary:} Overall, all these changes can be
subsumed in the kernel's code paths without requiring any changes to
the memory management API exposed to programmers. Moreover, every
single memory management function provided by stock Linux (e.g., {\it
  malloc}, {\it mmap}, {\it remap}, etc.) maintains the same
functionality to the end-user. Therefore, partitioning and data-PTE co-location is
achieved completely transparently to application developers and/or
system administrators.

\begin{algorithm}
  \caption{Physical page allocation scheme.}\label{alg:private}
  \begin{algorithmic}[1]
    \Procedure{alloc\_pages\_vma}{\textbf{addr\_t} vaddr}
    \State $\textit{mem\_set\_id} \gets \verb|MEM_SET_INDEX_HASH|(\mbox{vaddr})$
    \State $\textit{page} \gets \verb|alloc_pages_node|(\mbox{vaddr}, \mbox{mem\_set\_id})$
    \State \textbf{return} \textit{page}
    \EndProcedure
  \end{algorithmic}
\end{algorithm}

\noindent {\bf Copy-on-Write Support.} SPARTA supports any Copy-On-Write (CoW) scenario of the fork system call. A write on a read-only
page of a forked process will create a new page copy in the same partition as of the original page. Given that a page
can be mapped anywhere within a partition, SPARTA can support millions of concurrent fork calls and CoW operations.
Note that in the unlikely scenario of CoW pressure on a given partition, the CoW would still be supported but would
incur more swapping compared to a non-partitioned system. Also note that set-associative VM~\cite{picorel:near-memory} cannot
support CoW, as its hardware-based page-table does not have the flexibility to support synonyms.

%% file: tex/methodology.tex
\section{Experimental Methodology}
\label{sec:methodology}

\begin{table}[]
\caption{Workload description.}
\label{table:workload}
\resizebox{\linewidth}{!}{
\begin{tabular}{|c|c|l|}
\hline
\textbf{Workload} & \textbf{Footprint} & \multicolumn{1}{c|}{\textbf{Description}}                           \\ \hline
BST-External      & 128GB              & External Binary Search Tree from ASCYLIB \cite{david:asynchronized} \\ \hline
BST-Internal      & 128GB              & Internal Binary Search Tree \cite{david:asynchronized}              \\ \hline
Hash Table        & 128GB              & Hash Table \cite{david:asynchronized}                               \\ \hline
Skip Lists        & 128GB              & Skip-list based data index \cite{david:asynchronized}               \\ \hline
MultiProgrammed   & 4x32GB             & A multiprogrammed mix of the above four workloads                  \\ \hline
RocksDB           & 16GB               & Store engine running Facebook benchmarks \cite{facebook:rocksdb}    \\ \hline
\end{tabular}
}
\end{table}

Like recent work on VM~\cite{basu:efficient, pham:increasing, pham:colt, bhattacharjee:large-reach, barr:spectlb, papadopoulou:prediction-based, saulsbury:recently-based}, we use real-hardware measurements, trace-driven functional simulation, analytical models, and Linux kernel prototypes.

\subsection{Workloads}
To study TLB behavior at very large datasets (128GB), we use data traversal applications from the ASCYLIB~\cite{david:asynchronized} suite. We focus on state-of-the-art implementations of data indexing structures, such as  external and internal binary trees, skip lists, and hash tables,  which are the core of many server workloads such as Memcached and RocksDB. We choose these workloads because of their minimal data locality and instruction-level parallelism, and the fact stress general-purpose CPUs and hence favor custom memory-side hardware~\cite{ haria:devirtualizing, picorel:near-memory, kocberber:meet, hsieh:accelerating}. For space reasons we present the results for four representative implementations. To fully exercise the memory-partitioning support of our modified linux kernel v4.10 (which we plan to open-source) in both in-memory and out-of-memory setups, we use a more complex server workload, RocksDB, which is a store engine runninng Facebook's benchmarks. We run RocksDB with more realistic dataset sizes (16GB).

\subsection{TLB studies}
To study the impact of memory size on  TLB performance (Figure~\ref{fig:pagewalks}), we perform measurements on real hardware using the \textit{perf} tool during 10 minute windows. To study the TLB capacity requirements, we collect memory traces for 128GB datasets using Pin~\cite{luk:pin}, with each trace containing 1B instructions, which is more than enough for the TLB sensitivity experiments up to a few thousand entries. We use the traces to probe a set-associative TLB structure while varying its size and organization, after validating the baseline TLB model against the measurements on real hardware. 

\subsection{Performance}
Full-system simulation for TLB misses and page faults is not practical, as these events occur less frequently than other micro-architectural events (e.g., branch mispredictions). Hence, we resort to the CPI models often used in VM research~\cite{papadopoulou:prediction-based, saulsbury:recently-based, bhattacharjee:shared,picorel:near-memory} to sketch performance gains. We model the accelerators as simple in-order cores; the model includes the load/store frequency for each workload, hit rates of caches and memory- and accelerator-side TLBs, and the corresponding hit and miss latencies.

%% file: tex/new_evaluation.tex
\section{Evaluation}\label{sec:evaluation}

In this section, we evaluate various aspects of SPARTA and compare them
with conventional translation.

\subsection{Linux Prototype}
\begin{figure}[h]
   \centering
    \begin{minipage}{0.45\textwidth}
   \includegraphics[width=1.0\textwidth]{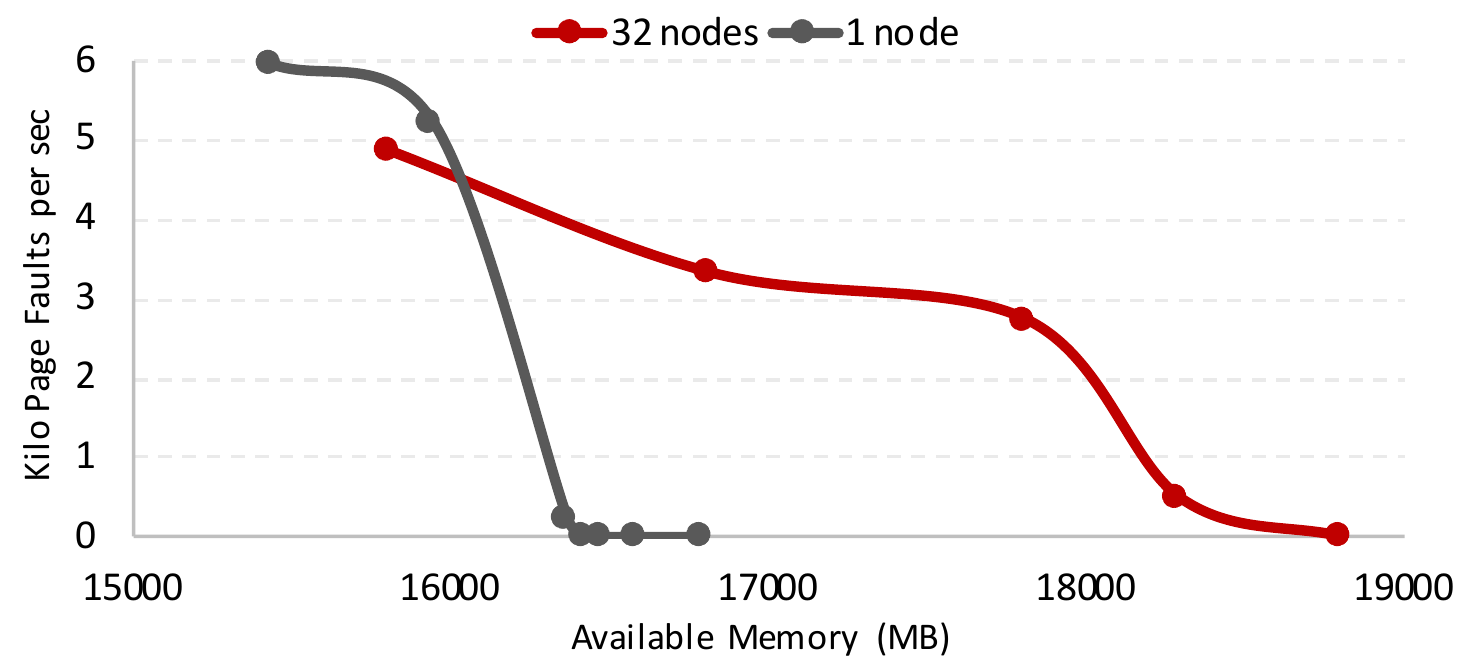}
   \caption{Page fault rate vs memory size for non-partitioned (1-node) and partitioned memory (32-nodes).}
   \label{fig:realhw}
    \end{minipage}
   \begin{minipage}{0.50\textwidth}
   \includegraphics[width=1.0\textwidth]{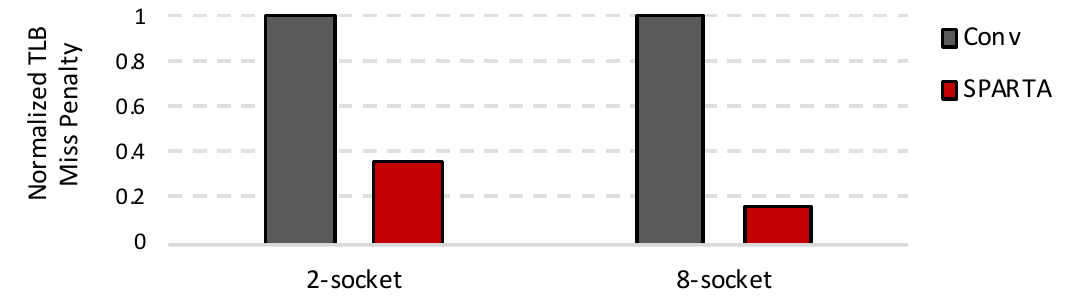}
   \caption{TLB miss penalty for conventional TLB organization and SPARTA.}
   \label{fig:penalty}
    \end{minipage}
\end{figure}

We successfully ran all server workloads on modified Linux with 32 memory partitions (nodes), utilizing above 90\% of available DRAM without a single page fault. Figure~\ref{fig:realhw} shows the page fault rate RocksDB tuned for a 16GB footprint, as we vary the available memory capacity in the critical range using the Linux boot-time $mem$ parameter. First, we note that our kernel properly deals with demand paging in the out-of-memory scenarios. Second, we see that the page fault rates in non-partitioned (1 node) and partitioned (32 nodes) case follow the same trend, except that the 32-partition setup requires about 1.5--2GB more memory to achieve the same page-fault rate. It is important to note that this phenomenon is not related to the tiny reduction in mapping flexibility caused by partitioning. Instead, this effect is the artifact of the prototype that relies on Linux NUMA nodes, which have small memory overheads per node and exhibit a degree of variability in their capacity.

\subsection{TLB Sensitivity}
\label{tlb sensitivity}

\begin{figure*}[t]
    \begin{minipage}{1 \textwidth}

    \centering
    \subfloat{
        \includegraphics[width=1 \textwidth,clip]{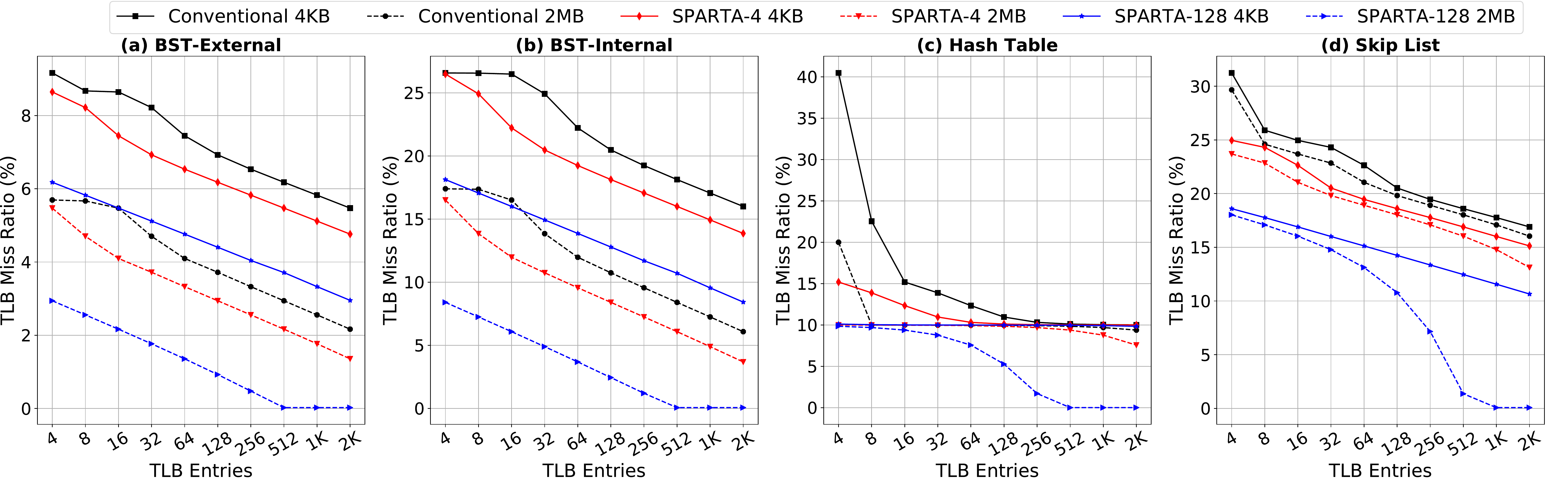}}
    \caption{TLB sensitivity study for 128GB working set for conventional translation with 4KB and 2MB pages and SPARTA.}
    \label{fig:miss_ratio_128GB}
        \end{minipage}

    \begin{minipage}{1 \textwidth}
        \centering
    \subfloat{
		\includegraphics[width=1 \textwidth,clip]{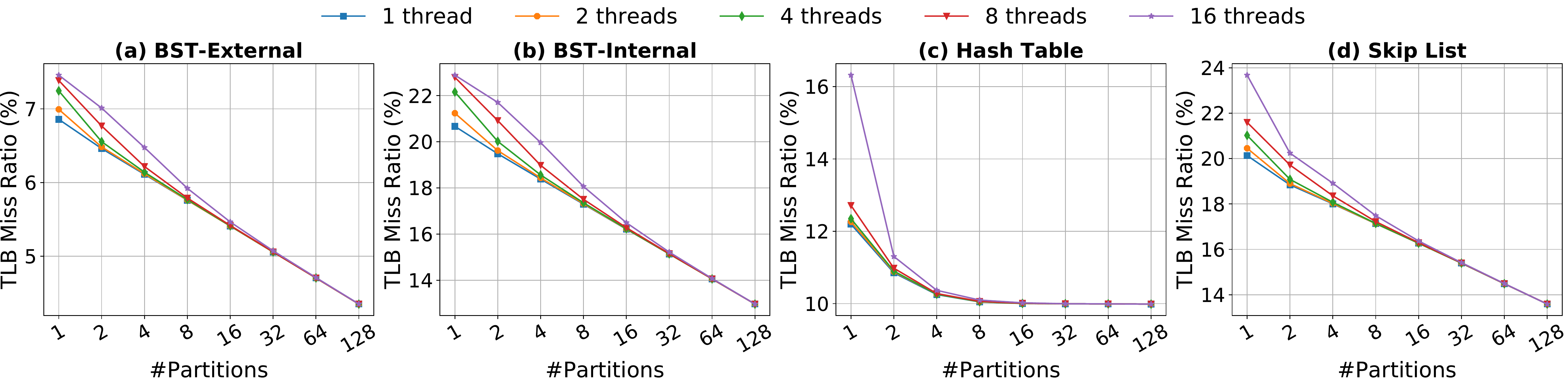}}

    \caption{The impact of thread contention on TLB miss rates.}
    \label{fig:contention}
    
    \end{minipage}
\end{figure*}

Figure~\ref{fig:miss_ratio_128GB} shows the TLB miss ratio for the four data traversal benchmarks: hash table, skip lists, internal and external binary search trees, as we vary the number of TLB entries. The working set size of all workloads is $128$GBs, and the TLB associativity is fixed to 4 ways. The black lines show the TLB miss ratio for the conventional translation (full black for 4KB and dotted black for 2MB pages). The full (dotted) red lines show TLB miss ratio for SPARTA with only four memory partitions (SPARTA-4) with 4KB (2MB) pages. It is important to note that the use of huge pages is orthogonal and fully compatible with our approach. Finally, the full and dotted blue lines show the TLB miss ratio for SPARTA with 128 memory partitions (DRIM-128) with 4KB and 2MB pages, respectively. 

For all workloads, the SPARTA-4 configuration with only 4 memory partitions (full red line) significantly outperforms the conventional translation with 4KB pages (full black line). On average, the memory-side TLBs require about 4x fewer TLB entries to match the performance of conventional execution-side TLBs. This efficiency of memory-side TLBs is attributed to their distributed nature: they cover only a fraction of the physical memory space. As expected, conventional TLBs with 2MB pages (black dotted line) perform strictly better compared to the TLBs with 4KB page sizes (full black line), often outperforming the SPARTA-4 configuration with 4KB pages (red full line) as well. However, for some workloads with poor spatial locality, such as skip lists, SPARTA-4 with 4KB pages is able to outperform even the conventional TLBs with 2MB pages with the same number of entries. 

Red dotted line shows TLB miss ratio for SPARTA-4 with 2MB pages. Not surprisingly, this configuration greatly outperforms the conventional designs with both 4KB and 2MB pages, as well as the SPARTA-4 configuration with 4KB pages. An interesting case is the SPARTA configuration with 128 partitions (SPARTA-128) and 4KB pages (full blue line). It outperforms all other configurations with 4KB pages, but also the configurations with 2MB pages for some workloads, such as skip lists. This shows that for workloads with low spatial locality, the benefits of further memory partitioning is much more significant than the impact of increasing the page size. Finally, the SPARTA-128 configuration with 2MB pages (blue dotted line) offers unmatched performance. Even with only 4 TLB entries this configuration is able to outperform the conventional configurations with 2048 TLB entries for most of the workloads and page sizes.

\subsection{Thread Contention}

Partitioning of the physical address space enables effective memory-side address translation with significantly
reduced TLB miss rates.  However, a major consequence of moving the
translation from the execution side to the memory partitions is that the
per-partition TLBs become shared by all threads in the system that are
accessing a given partition. A large system with many cores and
accelerators could have hundreds of threads, which can create two
potential problems related to TLB contention, which do not exist in
conventional TLBs. The first problem is the limited TLB bandwidth, and
the second problem is the interference among threads.

The first problem arises when TLBs cannot serve the volume of accesses generated by all threads. Fortunately, this problem does not occur in SPARTA. The reason is that even conventional TLBs are designed to sustain the L1 cache bandwidth of a single core, which is in the order of 100s of GB/s. In the case of SPARTA, the TLB bandwidth needs to be adequate to sustain the memory bandwidth of the local DRAM partition, which is an order of magnitude less. Furthermore, the size of memory partitions could be arbitrarily reduced. The second problem arises when different threads compete for the limited capacity of the shared per-partition TLB. To quantify this problem, we measure the TLB miss rates in SPARTA while varying the number of threads and the number of memory partitions. The results are shown in Figure~\ref{fig:contention}. In this experiment, the memory footprint of all applications is 128GB, the size of TLBs is 128 entries, and their associativity is fixed to four. The first thing to notice is that, in the single-threaded case, the TLB miss rates drop dramatically with the number of partitions as a result of physical address space partitioning, corroborating our real hardware experiments in Figure~\ref{fig:pagewalks}. The next thing to notice is that, in the case of a single memory partition, most applications exhibit a noticeable increase in TLB miss rates as we add more threads, as many threads contend for the capacity of a single shared TLB that covers the entire memory (a similar effect could be observed with hyper-threading, when multiple threads share the same TLB). However, once we start increasing the number of memory partitions, the impact of thread contention virtually disappears, showing that the benefits of partitioning significantly outweighs the impact of contention. This becomes most obvious when comparing (1 partition, 1 thread) and (16 partitions, 16 threads) data  points, because they have the same number of aggregate TLB entries per thread and the reduction in miss rates in entirely contributed to SPARTA's organization. Note that 512-entry TLBs with 2MB pages entirely cover a 1GB partition, and thus SPARTA-128 is guaranteed to eliminate TLB misses (except for Skip Lists, which has a footprint slightly above 128GB). Note that Figure~\ref{fig:contention}d is different from the rest, as it further illustrates the impact of multi-programming, described below. 

\subsubsection{Impact of Multiprogramming}
\begin{figure}[t]
   \centering
    
   \begin{minipage}{0.40\textwidth}
   \includegraphics[width=1.0\textwidth]{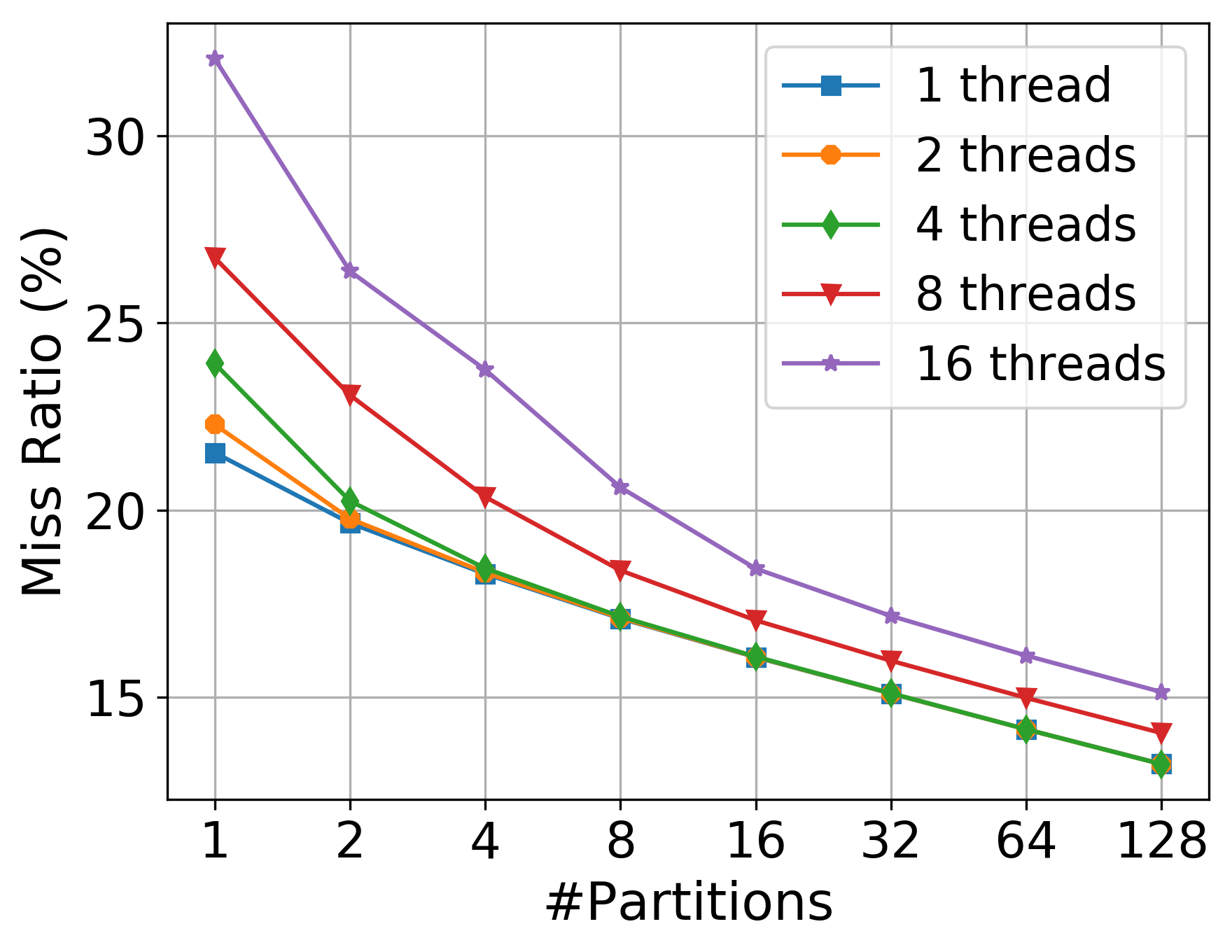}
   \caption{Impact of \textbf{Multiprogramming}. Blue, Orange and Green lines are all BST-E. Red line constitutes BST-E, BST-I and Hash Table, and for the Purpule line, we have added Skip List to rest of the workloads and measured the Miss Ratio.}
   \label{fig:multi-prog}
    \end{minipage}
    \begin{minipage}{0.55\textwidth}
   \includegraphics[width=1.0\textwidth]{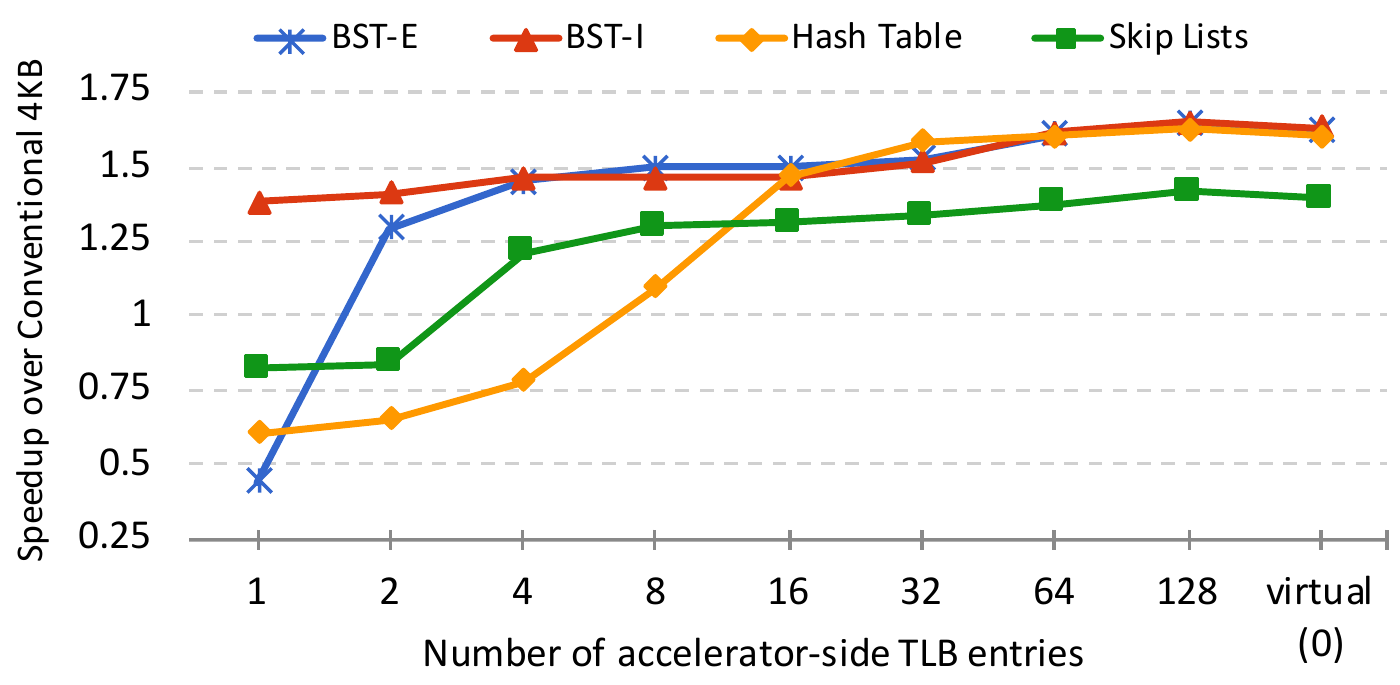}
   \caption{Impact of accelerator-side TLB capacity on SPARTA performance with physical caches. The rightmost datapoint reflects SPARTA with virtual caches and no accelerator-side TLBs.}
   \label{fig:speedup}
    \end{minipage}
\end{figure}

In Figure~\ref{fig:multi-prog}, we measure TLB miss rate starting with one thread of the BST-E application, and gradually increase the number of threads to 4 to study the impact of multithreading. Then we add 4 threads of another application (Hash Table) and plot the TLB miss ratio observed by the BST-E application (red line, 8 threads). Then we add the remaining two applications (BST-I and Skip List), four threads each, and observe the impact on BST-E Lists (purple line, 16 threads). The contention among the first four threads is minimal, as they belong to the same application and share the dataset, which SPARTA leverages by avoiding redundant caching of TLB entries. However, as we add unrelated applications, the contention in TLBs naturally increses. Despite the increased contention, SPARTA manages to significantly reduce the TLB miss ratio through partitioning. While we show the impact of multi-programming only on BST-E, we have confirmed the same trends for other workloads.

\subsection{TLB Miss Penalty}
SPARTA does not only improve the area-efficiency of TLBs but also reduces the TLB miss penalty. Figure~\ref{fig:penalty} shows the average TLB miss penalty for two systems, a two-socket machine and a large-memory eight-socket one. The numbers are normalized to the conventional design. We decided to be highly conservative and assume that in the baseline case 1) the accelerators have a private MMU instead of relying on a shared IOMMU, thus excluding the IOMMU communication latency, and 2) the accelerator's MMU includes perfect MMU caches, and therefore page walks require only one memory reference. This also helps us isolate our primary TLB miss penalty benefits stemming from translation/data fetch overlap from the additional benefits our particular page table implementation provides, since in our case, with an inverted page table ($1/4x$ load factor), we are able to guarantee one memory reference per page walk, whereas the conventional four-level page table often requires multiple memory accesses despite having a complex MMU cache hierarchy.

The miss penalty in the conventional case is fully exposed and comprises of a DRAM access and two network traversals (request and response), as explained in Figure~\ref{fig:timeline}b. For SPARTA, in contrast, the time it takes to locate the partition, traverse the NoC and off-chip interconnects is overlapped with the data path (Figure~\ref{fig:timeline}d). As a result, the TLB miss latency only includes DRAM latency. As Figure~\ref{fig:penalty} shows, the larger the memory system, the larger the contribution of the network time in the average page walk time, and hence the larger the SPARTA's reduction in page walk time with respect to the conventional translation. This makes SPARTA particularly attractive for future memory systems, which are, due to the end of Moore's law, likely to be physically larger and distributed in a NUMA fashion.

\subsection{Performance}

\begin{figure}[t]
	\begin{minipage}{1\textwidth}

	\centering
	\includegraphics[width=1\textwidth]{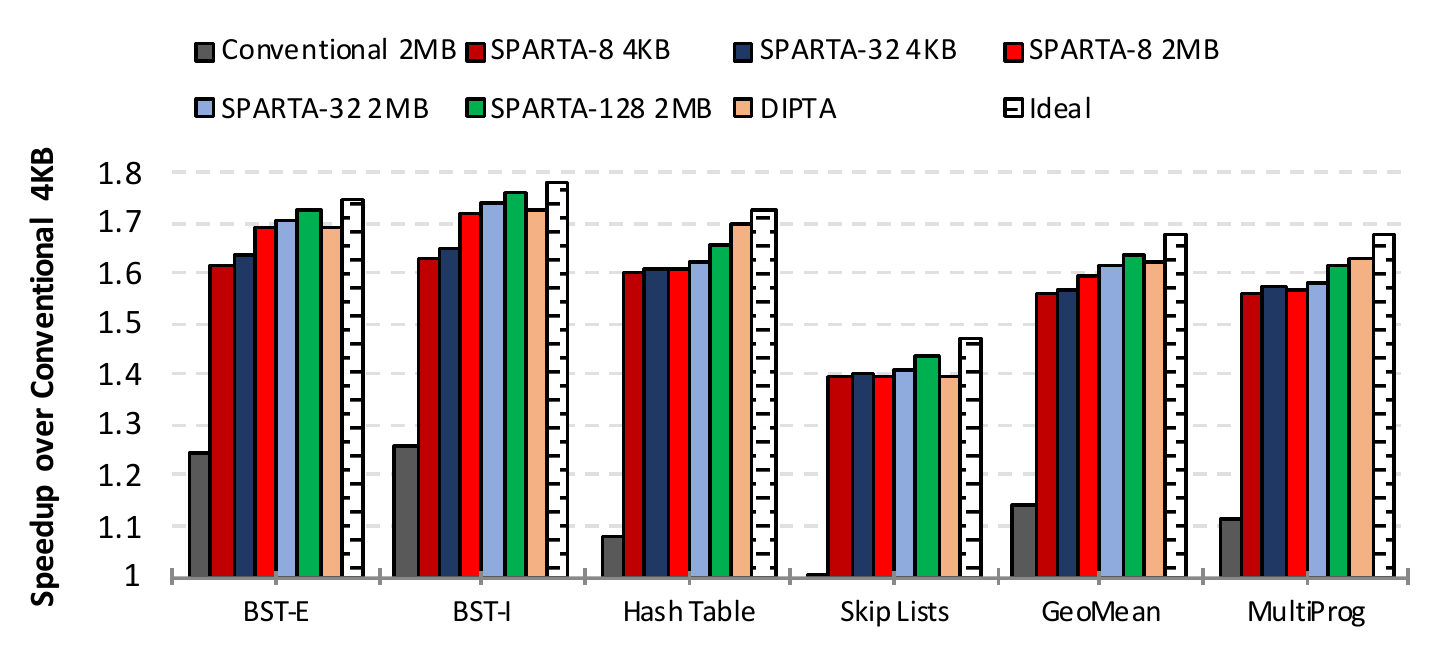}
	\caption{Performance improvement of SPARTA with 4 and 32 partitions over the baseline.}
	\label{fig:performance}
	
	\end{minipage}
\end{figure}

Figure~\ref{fig:performance} shows the performance improvement of SPARTA-8 (one
partition per socket) and SPARTA-32 (one partition per memory channel) for an 8-socket 
machine with 128GB of memory (4 channels/socket, as in Figure~\ref{fig:overview}). 
We consider both 4KB and 2MB page sizes for both SPARTA and conventional translation
and normalize the results to conventional 4KB. Each accelerator includes a 16KB 4-way 
associative virtual cache. The baseline systems have accelerator-side TLBs and perfect MMU caches, whereas in 
SPARTA accelerators do not have any translation hardware. All TLBs have 128 entries. 
The ideal system has no zero overhead.

The conventional 2MB design provides an average speedup of only 14\%. In contrast, even 
the weakest SPARTA configuration (SPARTA-8 4KB) outperforms the conventional 4KB design 
by 1.4x on average, outperforming the conventional 2MB design for every workload individually. 
When comparing SPARTA-8 2MB and SPARTA-32 4KB designs, increasing the page size has more 
impact than further partitioning more all workloads other than SkipLists, which has low spatial locality. 
On average, SPARTA-32 improves performance by 1.57x for 4KB pages (1.62x for 2MB), which is 
within 93.7\% of the ideal performance (96.2\% for 2MB). We also show the most powerfull reasonable design point, SPARTA-128,
which achieves 98\% of the ideal performance with 2MB pages.

\subsection{Impact of Cache Designs}
Figure~\ref{fig:speedup} shows the performance improvement of SPARTA-8 (one partition per socket) 
for an 8-socket machine with 128GB of memory over the baseline system with 4KB pages. 
Each accelerator includes a 16KB 4-way associative cache, organized  as either physical or virtual cache. 
In the baseline system, every accelerator has a 128-entry TLB and perfect MMU caches, while in SPARTA 
we vary the number of accelerator-side TLB entries from 1--128 for physical caches (with no MMU caches), and use no 
translation hardware for virtual caches. Memory-side TLBs have 128 entries and no MMU caches. The first thing to note 
is that SPARTA with physical caches needs only 8 accelerator-side TLB entries to outperform the baseline with 128 entries 
for all applications. Additional TLB capacity beyond 8 entries provides diminishing returns, as it unnecessarily 
extends the TLB reach beyond the cache. SPARTA sees little benefit from that as it largely overlaps the handling of cache and TLB misses. 
Note that virtual caches without TLBs offer slightly lower performance compared to 
physical caches with large TLBs. While data residing in virtual caches needs no translation, for the fraction of 
data that is not in the cache but is covered by a large TLB, physical caches see no 
translation overhead whereas SPARTA may not always succeed to fully overlap translation and fetch.

\subsection{Comparisson with Set-Associative VM}
As explained in Section~\ref{sec:os}, set-associative VM systems cannot implement their page tables in software. Consequently, DIPTA~\cite{picorel:near-memory} implements its page table in hardware, storing translations either in SRAM or DRAM. For a 128GB memory system we study, we find that SRAM-based DIPTA requires 256MB of SRAM storage, or 32-128MB per socket, which is out of reach for today's processors. While DRAM-based DIPTA appears to be a better alternative, it requires custom-designed 3D-stacked DRAM. Furthermore, its 1.5\% DRAM overhead per each way of associativity is non-negligible. While 4-way associative DIPTA does not incur significant page-faults for single-process workloads, we find that our multi-programmed workload needs 16 ways to avoid significant slowdowns due to page faults. As a result, DRAM-based DIPTA must sacrifice 24\% of DRAM capacity for the page table to run workloads with a moderate degree of multiprogramming. Regardless of the implementation, DIPTA requires way prediction to hide the latency of page-table lookups.

Figure~\ref{fig:performance} also shows the speedup DIPTA achieves relative to the baseline. Because we cannot reliably estimate the latency of probing a 256MB SRAM structure, we assume an idealized DRAM-based DIPTA that does not incur any DRAM overhead. As Figure~\ref{fig:performance} shows, DIPTA falls short of not only the ideal translation, but also of several SPARTA configurations due to way mispredictions. For many workloads, even weaker SPARTA configurations outperform DIPTA, as the hit ratio of SPARTA's memory-side TLBs are higher than the accuracy of DIPTA's way prediction. The only exception is Hash Table, which enjoys a very high (>90\%) way prediction accuracy, and MultiProgrammed workload, where SPARTA sees less TLB sharing benefits. Regardless of the performance results, we find that DIPTA is not a practical design due to its prohibitive overhead and its lack of support for the key VM features.

%% file: tex/relatedwork.tex
\section{Discussion \& Related Work}
\label{sec:relatedwork}

\noindent\textbf{Improving TLB performance.} The techniques described in Section~\ref{sec:background} improve TLB reach but others reduce TLB miss penalty \cite{bhattacharjee:tempo}. Processors store page table entries in data caches to accelerate page walks~\cite{intel:architectures} and many also use page table caches (e.g., TSBs in SPARC~\cite{sun:ultrasparc}). MMU caches are employed to skip walking intermediate page table levels~\cite{bhattacharjee:large-reach, barr:translation, margaritov2019prefetched}. Other techniques translate speculatively~\cite{barr:spectlb} or prefetch TLB entries~\cite{bhattacharjee:characterizing}. LegoOS~\cite{shan:legoos} recently argued that in disaggregated memory systems address translation should be delegated to memory, and suggested using virtual caches on the execution side. All these techniques are orthogonal to SPARTA.

\vspace{2mm}
\noindent\textbf{Relationship with NUMA.} Modern OSes allow users to bind application threads to specific NUMA nodes, as well as to allocate memory from a given NUMA node. Such primitives can be used to collocate computation and data in the same NUMA node, reducing the data fetch latency. Unfortunately, the OSes do not provide a way to control the placement of the page tables. Consequently, such setups suffer from high translation overhead, as data can be local while translations are often remote \cite{achermann2019mitosis}. We hope that SPARTA's collocation of data and page tables will inspire future OSes to incorporate intelligent NUMA-aware placement/replication/partitioning of conventional page tables to benefit big-memory systems.

\vspace{2mm}
\noindent\textbf{Relationship with page coloring.} Like SPARTA, page coloring also influences physical page placement, but with a purpose of improving cache performance~\cite{chiueh:eliminating, gustafson:ibm, jouppi:architectural}. It does not target TLBs, leading to important operational differences. Namely, page coloring is largely used as a performance optimization seeking to reduce cache conflicts, whereas SPARTA {\it requires} a specific mapping function to be maintained between virtual and physical addresses to {\it correctly} route requests to a memory partition and its TLB. Multiple different virtual-to-physical address mapping functions can be used to realize coloring, but SPARTA mandates exactly one for correct operation. These differences mean that OS implementations of SPARTA are entirely different than page coloring. However, SPARTA could interfere, positively or negatively, with page coloring, and future work could study how to integrate page coloring {\it with} SPARTA. However, we expect that the tiny restrictions in virtual-to-physical mapping that SPARTA imposes would not significantly affect OS opportunity to apply page coloring on top of SPARTA. 





%% file: tex/conclusion.tex
\section{Conclusion}
\label{sec:conclusion}

In this work we argue that splitting the task of address translation between accelerators and memory results in significant performance and efficiency gains. We show that overlapping the memory-side translation with data fetch allows for practical and efficient accelerator-side TLBs. We further show that only minor restrictions in virtual-to-physical address mapping are sufficient to almost entirely overlap address translation and data fetch and achieve the same performance benefits as the extremely restrictive solutions, without sacrificing key VM features. By allowing a virtual address to identify a memory chip and partition uniquely, memory can be accessed as soon as the virtual address is known, while a memory-side TLB translates and fetches the data, overlapping both operations almost entirely. By further co-locating data and PTEs in the same partition, memory-side page-walks become localized and fast. By implementing this concept in stock Linux prototype, we show that partitioned memory with localized page-tables can be easily implemented using the existing Linux code paths.

\newpage